\documentclass[a4paper,aps,prl,floatfix,onecolumn,nofootinbib]{revtex4-1}

\usepackage{bm}			    
\usepackage{amsmath} 
\usepackage{amssymb}
\usepackage{latexsym}
\usepackage{amsfonts}
\usepackage{color}
\usepackage{soul} 
\usepackage{graphicx}

\newcommand{\ket}[1]{\vert{#1}\rangle} 
\newcommand{\bra}[1]{\langle{#1}\vert} 
 
\newcommand{\op}[2]{\ket{#1}\!\bra{#2}}
\newcommand{\mean}[1]{\langle #1 \rangle}

\DeclareMathOperator{\Tr}{Tr}

\renewcommand{\vec}[1]{\mathbf{#1}}

\newcommand{\beq}{\begin{equation}}
\newcommand{\eeq}{\end{equation}}

\newcommand{\be}{\begin{equation}}
\newcommand{\ee}{\end{equation}}
\newcommand{\esp}[1]{\left< #1 \right>}
\newcommand{\cor}[1]{\left[ #1 \right]}
\newcommand{\pare}[1]{\left( #1 \right)}
\newcommand{\key}[1]{\left\{ #1 \right\}}
\newcommand{\ben}{\begin{eqnarray}}
\newcommand{\een}{\end{eqnarray}}

\begin{document}

\title{Quantum transport in $d$-dimensional lattices}


\author{Daniel Manzano$^{1,2}$}
\email{manzano@onsager.ugr.es}
\author{Chern Chuang$^{1}$}
\author{Jianshu Cao$^{1}$}
\email{jianshu@mit.edu}

\affiliation{$^{1}$ Massachusetts Institute of Technology,
Chemistry Department. Cambridge, Massachusetts 02139, USA  }
\affiliation{$^{2}$  Singapore University of Technology and Design, Engineering Product Development.
8 Somapah Road, Singapore 487372}
\date{\today}



\begin{abstract}
We show that both fermionic and bosonic uniform $d$-dimensional lattices can be reduced to a set of independent one-dimensional chains. This reduction leads to the expression for ballistic energy fluxes in uniform fermionic and bosonic lattices. By the use of the Jordan-Wigner transformation we can extend our analysis to spin lattices, proving the coexistence of both ballistic and non-ballistic subspaces in any dimension and for any system size. We then relate the nature of transport to the number of excitations in the homogeneous spin lattice, indicating that a single excitation always propagates  ballistically and that the non-ballistic behavior of uniform spin lattices is a consequence of the interaction between different excitations. 
\end{abstract}

\maketitle

\section{Introduction}
Due to the rapid advance in nano-technologies the study of non-equilibrium transport phenomena in quantum systems, including charge and heat transport, has become a major field of study.  Furthermore, transport properties in spin and harmonic oscillators lattices are important because they model many realistic physical systems. Spin lattices and ladders can be experimentally realized by different methods \cite{bachelor:ap07,hess:epj07,hild:prl14}. Harmonic oscillators lattices describe a plethora of different systems including optical cavities arrays \cite{hartmann:lpr08}, trapped ions \cite{bermudez:prl13}, and phonons in cubic crystal lattices \cite{krauth_06,dhar:pre12}. Finally, fermionic and bosonic atoms transport can be studied in optical lattices showing both ballistic and diffusive regimes \cite{Bloch-f,bloch-b} 

The role of dimensionality is crucial in both quantum and classical systems \cite{dhar:ap08}. One-dimensional systems composed by spins  \cite{znidaric:jsm10,manzano:pre12} and harmonic oscillators \cite{asadian:pre13} have been previously analysed by the use of Markovian master equations. These analyses show that these systems are ballistic when the transport takes place coherently and become diffusive if a dephasing channel is locally coupled. This is a typical behavior of one-dimensional homogeneous systems in the presence of noise, while disordered systems possess additional features \cite{moix:njp13}. The same behavior is found in harmonic lattices of arbitrary dimension \cite{asadian:pre13}. However, in multidimensional spin lattices the results are different, and an analytical solution is lacking. The simplest two dimension topology, a ladder, has been analysed by \v{Z}nidari\v{c} \cite{znidaric:prl13}, showing that quantum transport is in general anomalous. In fact, this system has a set of ballistic subspaces, making it possible to create ballistic transport by the design of non-local baths operators and the initial state preparation. It has also been demonstrated experimentally with ultracold atom in optical lattices that spin systems of one and two dimension behave very differently\cite{hild:prl14}.

Another canonical problem regarding multidimensional transport is quantum walk (QW). It is known that a QW in a one dimensional system propagates faster than classical random walks \cite{venegasandraca:qip12}.  In multidimensional systems, it has been proved by Kempe \cite{kempe:ptrf05} that a discrete  QW in a hypercube hits from corner to corner exponentially faster than a classical random walk. This result has been extended to distorted hypercubes \cite{krovi:pra06} and to hypercubes embedded in more complex graphs \cite{makmal:pra14}. Finally, in hypercubic lattices some searching algorithms based on QW have been developed showing a speed-up against their classical counterparts  \cite{patel:pra10,childs:pra14}. These results suggest that qubit lattices with only one excitation behave in a superdiffusive way.

Regarding the transport in quantum lattices there are still some open questions:

\begin{itemize}
\item What is the role of the fermionic and bosonic nature of the system in the transport properties?
\item Why spins and harmonic oscillator systems behave differently in lattices with dimension $d>1$?
\item What is the origin of the ballistic and non-ballistic subspaces found in spin ladders?
\item How does the number of excitations in a spin lattice affect transport properties?
\end{itemize}

In this paper we study the general case of quantum transport in nonequilibrium lattices composed of fermionic or bosonic sites. We show that these lattices with an arbitrary dimension can be decomposed into independent one-dimensional normal modes. This confirms that both fermionic and bosonic lattices are ballistic in any dimension and allows us to derive explicit expressions for energy fluxes of uniform fermionic and bosonic lattices. We then use this method to analyse spin $d$-dimensional lattices showing that ballistic spaces exist in any dimension and size of the system. Finally, we discuss the implications of this result in hypercubic lattice quantum walks and discuss the role of the number of excitations in  transport properties.

\section{Lattice model}
The systems we study are fermionic and bosonic $d$-dimension lattices. In order to analyse the features of the transport we couple the system to incoherent thermal baths that drive it out of equilibrium. Due to the effects of the baths, the system evolves to a time-independent steady state. If the temperatures of the baths are different, there is a finite energy current flowing through the system. The dependence of the current on the system size indicates the nature of the transport. We consider transport to be ballistic if the current is independent of the system size \cite{manzano:pre12,asadian:pre13}.

\begin{figure}
\hspace{-1cm}
\includegraphics[scale=0.36]{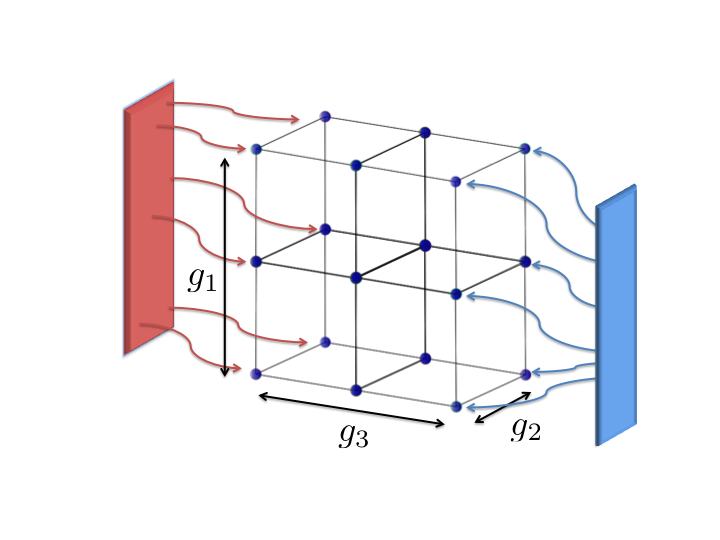}
\caption{Sketch of a $d=3$ lattice with $L_1=3$, $L_2=2$, and $L_3=3$. Local thermal baths apply at the terminal sites of dimension $3$.}
\label{fig:system}
\end{figure}

In figure \ref{fig:system} a $d=3$ lattice with local thermal baths at the end of dimension $3$ is displayed. Each site of the lattice is defined by $d$ indices $l_i=1,\dots,L_i$, where $L_i$ is the size of the system in dimension $i$. We define the site vector $\vec{l}=\pare{l_1,\dots,l_d}$ and the size vector $\vec{L}= \pare{L_1,\dots,L_d}$. The lattice is connected to thermal baths at the ends of dimension $d$, making this the direction where the energy transfer takes place. Periodic boundary conditions are applied to all the remaining dimensions. We also define the reduced vectors to represent all the dimensions but the last one, $\vec{l}_r=\pare{l_1,\dots,l_{d-1}}$ and  $\vec{L}_r=\pare{L_1,\dots,L_{d-1}}$. 

The full Hamiltonian of our system can be decomposed into the free Hamiltonian, which depends only on the occupation number of each site, and the hopping Hamiltonian in each dimension, which is responsible for transport. We can write the total Hamiltonian as $H=H_{\text{free}}+\sum_{i=1}^d H_i$ (we take  $\hbar=1$ throughout the paper)   

\ben
H_{\text{free}}&=&  \omega \sum_{\vec{l}}  a^\dagger_{\vec{l}} a^{\phantom{\dagger}}_{\vec{l}}  \nonumber\\
H_i&=& g_i \sum_{\vec{l}} \pare{ a^\dagger_{l_1,\cdots,l_i,\cdots,l_d} a^{\phantom{\dagger}}_{l_1,\cdots,l_i+1,\cdots,l_d}  + \text{H.c.} },
\label{eq:Hd}
\een
were $\omega$ is the frequency of each site, and $g_i$ represents the coupling strength in dimension $i$. The operators $a$ and $a^\dagger$ are ladder operators that can be either fermionic $(f^{\phantom{\dagger}},f^\dagger)$ or bosonic $(b^{\phantom{\dagger}},b^\dagger)$, obeying anticommutation and commutation relations respectively.  In the fermionic lattice, as there are no internal degrees of freedom each site can contain at most one fermion ($f_i^\dagger f_i^\dagger=0$). On the contrary, the bosonic lattice can contain an arbitrary number of bosons per lattice site. 

The system is driven out of equilibrium by bosonic thermal baths locally coupled to the end of the system in dimension $d$. The overall dynamics of the system is described by a Markovian master equation \cite{breuer_02}.

\be
\dot{\rho}=-i\cor{H,\rho} + \mathcal{L}_1 \rho + \mathcal{L}_{L_d} \rho 
\label{eq:ME}
\ee
Here, each of the reservoirs is modelled by a Lindblad super-operator

\ben
\mathcal{L}_{1/L_d} \rho &=&  \sum_{\vec{l}_r}  \Gamma_{1/L_d} n_{1/L_d} 
\pare{ a^\dagger_{l_1,l_2,\dots,1/L_d} \rho \; a^{\phantom{\dagger}}_{l_1,l_2,\dots,1/L_d} 
 -\frac{1}{2} \key{a^{\phantom{\dagger}}_{l_1,l_2,\dots,1/L_d}a^\dagger_{l_1,l_2,\dots,1/L_d} ,\rho }} \nonumber\\
&+&  \sum_{\vec{l}_r} \Gamma_{1/L_d} \pare{ n_{1/L_d}+1}
\pare{ a^{\phantom{\dagger}}_{l_1,l_2,\dots,1/L_d} \rho \; a^\dagger_{l_1,l_2,\dots,1/L_d} 
-\frac{1}{2} \key{a^\dagger_{l_1,l_2,\dots,1/L_d} a^{\phantom{\dagger}}_{l_1,l_2,\dots,1/L_d} ,\rho } }   
\label{eq:lindblad}
\een
where $\mathcal{L}_{1}$ acts at the beginning of dimension $d$ and $\mathcal{L}_{L_d}$ models the bath attached at the end of this dimension. The first term in $\mathcal{L}_i$ accounts for emission into the reservoir, and the second term for absorption, $\Gamma$ is the interaction rate,  and $n_i$ is the mean excitation number at the resonance frequency of the reservoir. For instance, in the case of a bosonic bath  at temperature $T_i$, $n_i=1/[\exp (\omega/T_i) - 1]$ (with Boltzmann's constant $k_B=1$). We perform the analysis using Bose-Einstein statistics, but the results are independent of the nature of the baths.

The validity of this master equation in nonequilibrium systems has been extensively analysed. In Ref. \cite{rivas:njp10} a system of two interacting harmonic oscillators coupled to thermal baths at different temperatures is studied both numerically and analytically, concluding that the master equation (\ref{eq:ME}) is valid in the limit of small intercoupling. Furthermore, in Ref. \cite{asadian:pre13} it is proved that in the case of bosonic lattices with two local baths at the same temperature the system thermalises in the limit of small intercoupling.

The energy transfer through the lattice can be quantified by the time derivative of the energy expectation value in the system

\be 
\dot{E}=\frac{d}{dt} \esp{H}=\Tr \pare{H\dot{\rho}}.
\label{eq:energy}
\ee
This expression should be zero at the steady state. Using the master equation (\ref{eq:ME}) we can decompose (\ref{eq:energy}) into 

\be
\dot{E} = \Tr\pare{H  \mathcal{L}_1 \rho} + \Tr\pare{H \mathcal{L}_2 \rho} := J_1+J_2=0,
\ee
where $J_1$ and $J_2$ are equal in magnitude but with opposite signs. Each of these terms refers to the mean energy interchanged with each thermal bath per unit of time. This allows us to define the energy flux through the system as  $J=|J_1|=|J_2|$. In regular systems the behaviour of the energy flux is proportional to the behaviour of the spin or excitation flux. In more complicated systems like networks the energy and excitations fluxes can behave differently \cite{manzano:po13}.

\section{Normal-mode decomposition} 

To calculate the energy transfer through an arbitrary lattice, we define normal modes associated with all the dimensions except for $d$, $q_\alpha^{n_\alpha}=\frac{2 \pi n_\alpha}{L_\alpha}$, where $\alpha=1,\dots,d-1$ and $n_\alpha= 1,\dots, L_\alpha$ \cite{roy:jsp08,nakazawa:ptps70}. Therefore, we define the mode vector as $\vec{q}=\pare{q_1^{n_1},\dots,q_{d-1}^{n_{d-1}}}$ and there are $N= \prod_{i=1}^{d-1} L_i$ normal modes. We transform the ladder operators to the normal mode basis via the following transformation

\ben
a_{l_1,\dots,l_d}&=&\frac{1}{\sqrt{N}}    \sum_{\vec{q}} a^{\phantom{\dagger}}_{l_d}\pare{\vec{q}}    \exp \cor{ 2\pi i \sum_{i=1} ^{d-1} \frac{q_\alpha^{n_{\alpha}} l_i}{2\pi} } \nonumber\\
a^\dagger_{l_1,\dots,l_d}&=&\frac{1}{\sqrt{N}} \sum_{\vec{q}}  a^\dagger_{l_d}\pare{\vec{q}}    \exp \cor{ -2\pi i \sum_{i=1} ^{d-1} \frac{q_\alpha^{n_{\alpha}} l_i}{2\pi} },
\label{eq:modetrans}
\een
where $a^{\phantom{\dagger}}_{l_d}$ and  $a^{\dagger}_{l_d}$ are ladder operators in the normal mode basis that fulfils the same commutation/anticommutation relations as the site ladder operators. Applying this transformation to the Hamiltonian (\ref{eq:Hd}) decomposes it into independent terms for each mode, $H=\sum_{\vec{q}} H(\vec{q}) $ with

\be
H(\vec{q}) =\pare{ \omega +2 \sum_{k=1}^{d-1} g_k \cos \pare{q_{\alpha}^{n_\alpha} }  } \sum_{l_d=1}^{L_d}  a^\dagger_{l_d}\pare{\vec{q}}  a^{\phantom{\dagger}}_{l_d}\pare{\vec{q}} 
+ g_d \sum_{l_d=1}^{L_d-1} \pare{ a^\dagger_{l_d}\pare{\vec{q}}  a^{\phantom{\dagger}}_{l_d+1}\pare{\vec{q}} + \text{H.c.}}.
\label{eq:H_modes}
\ee

Also the Lindblad super-operators (\ref{eq:lindblad}) decompose into independent terms for each normal mode

\begin{widetext}
\ben
\mathcal{L}_{1/L_d} \pare{\vec{q}} \rho &=& 
\Gamma_{1/L_d}  n_{1/L_d} 
\pare{ a^{\phantom{\dagger}}_{1/L_d} \pare{\vec{q}} \rho a_{1/L_d}^\dagger \pare{\vec{q}} - \frac{1}{2} \key{a_{1/L_d}^{\dagger}\pare{\vec{q}} a^{\phantom{\dagger}}_{1/L_d}\pare{\vec{q}} ,\rho } }  \nonumber\\
&+& \Gamma_{1/L_d}  \pare{ n_{1/L_d}+1} \pare{ a_{1/L_d}^\dagger \pare{\vec{q}} \rho a^{\phantom{\dagger}}_{1/L_d}\pare{\vec{q}}  \right.
\left.  -\frac{1}{2} \key{ a^{\phantom{\dagger}}_{1/L_d}\pare{\vec{q}}  a_{1/L_d}^\dagger \pare{\vec{q}} ,\rho}}. 
\label{eq:lindblad-mode}
\een
\end{widetext}

The master equation  (\ref{eq:ME}) unravels into a set of $N$ independent equations, each of them corresponding to one of the modes, which behaves as bosonic or fermionic one-dimensional system. Each of these equations is equivalent to the equation of a one dimension system with the on-site energies shifted by $2 \sum_{k=1}^{d-1} g_k  \cos \pare{q_{\alpha}^{n_\alpha} }$. 

Hence, the problem of calculating the energy flux through a $d$-dimensional lattice can be reduced to calculating the energy current in a one-dimensional chain. The heat transfer through one-dimensional chains composed by harmonic oscillator has already been solved in Ref. \cite{asadian:pre13}. The one-dimensional fermionic chain is solved in the Appendix. The analytical expressions for the energy flux for each mode is

\ben
J_\text{F}\pare{\vec{q}} &=& \pare{\omega +2 \sum_{k=1}^{d-1} g_k \cos \pare{q_{\alpha}^{n_\alpha} }}
\frac{4g_d^2 \gamma_1\gamma_{L_d} (s_1-s_{L_d})} {\pare{\gamma_1+\gamma_{L_d}} \pare{4g_d^2+\gamma_1 \gamma_{L_d}} }
\nonumber\\
J_\text{B}\pare{\vec{q}} &=& \pare{\omega +2 \sum_{k=1}^{d-1} g_k \cos \pare{q_{\alpha}^{n_\alpha} } } 
 \frac{4 g_d^2 \Gamma_{1} \Gamma_{L_d}  (n_1-n_{L_d})}{\pare{\Gamma_1+\Gamma_{L_d}} \pare{4 g^2_d+\Gamma_1 \Gamma_{L_d}}},
\label{eq:current}
\een
where we have introduced the notation $\gamma_i=\Gamma_i(2 n_i + 1)$ and $s_i=n_i/(2n_i+1)$ in the fermionic case.  The total energy flux can be calculated by summing up the flux of each channel, $J_{\text{total}}=N J_{1d}$ where $J_{1d}$ is the energy flux of a uniform one-dimension system with on-site energies $\omega$. 
 
The lattice size in the energy transfer dimension plays no role in the amount of energy transfer, as it does not appear in (\ref{eq:current}). This result shows that both fermionic and bosonic transport in the absence of noise or disorder are ballistic with arbitrary size and in any dimension. As our model coincides with the Hubbard model with no interactions (U=0), this result explains why in the non-interacting regime there is ballistic transport in optical lattices both with fermionic and bosonic atoms \cite{Bloch-f,bloch-b}.

\begin{figure}[h]
\includegraphics[scale=0.7]{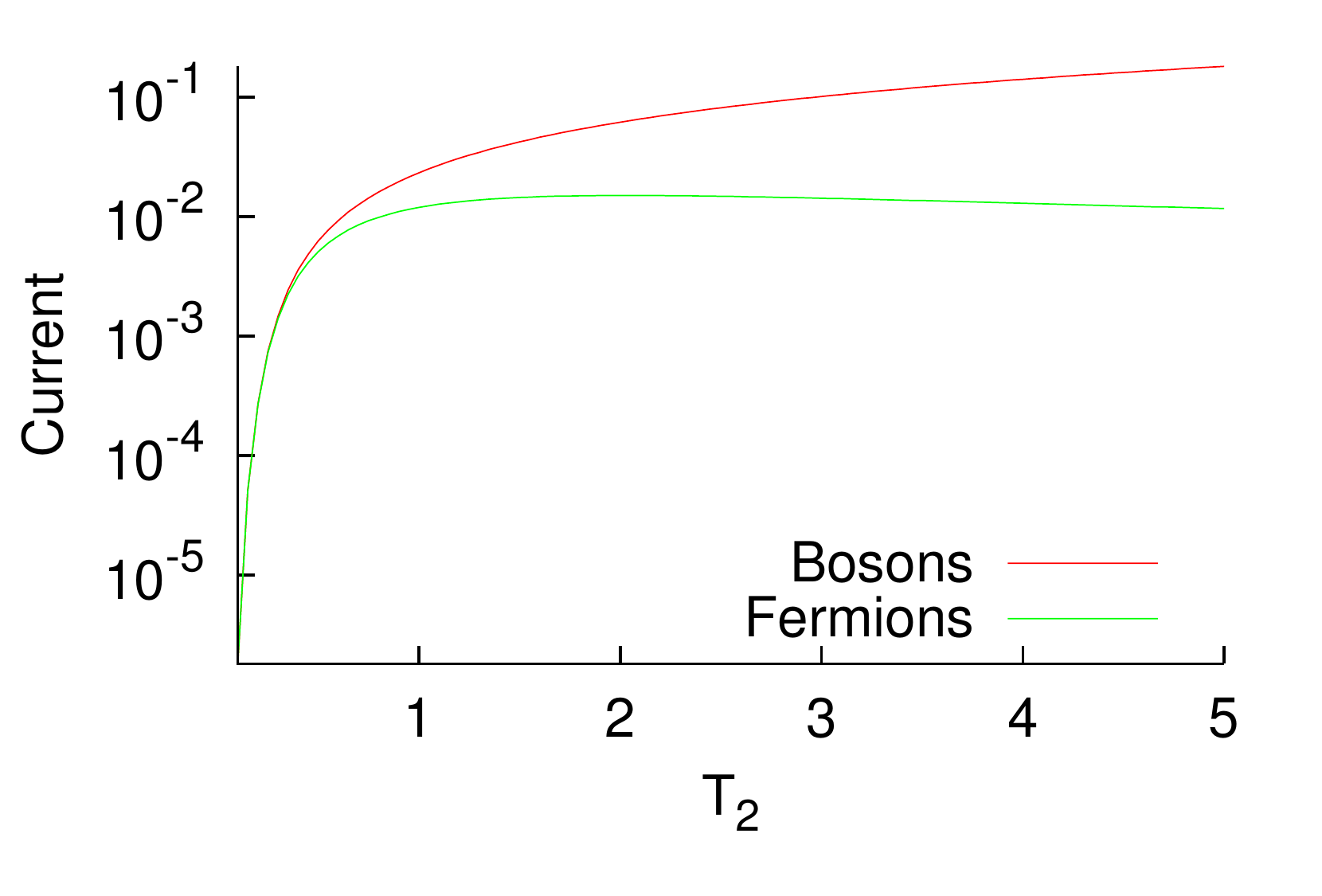}
\caption{ Energy current (logarithmic scale) for a fermionic and bosonic system as a function of the temperature of the hot bath. $\Gamma_1=\Gamma_2=0.01,\; g=0.01,\;\omega=10, T_1=0.001$, $d=1$.}
\label{fig:current}
\end{figure}

Even though both fermionic and bosonic transport are ballistic, their dependences on parameters such as the temperatures are different. In the limit of zero temperature the two converge as $\lim_{T_i\to 0} \gamma_i=\Gamma_i$ and $\lim_{T_i\to 0} s_i=n_i$. But the situation at finite temperature is very different. The energy currents of one dimensional fermionic and bosonic systems are displayed in figure (\ref{fig:current}) as functions of the temperature of the hot bath (the cold bath is kept constant at a very low temperature). For small values of the hot bath temperature the currents are very similar in magnitude. At higher temperatures the bosonic system has a higher energy current than the fermionic one. This is a consequence of the Pauli exclusion principle that limits the heat capacity of the fermionic system.  In the limit of very high temperature of the hot bath the current of the fermionic lattice decreases as the temperature increases, similar to spin systems \cite{manzano:pre12}.

Finally, we need to point out that there are also non-ballistic systems that can be decomposed into non-interacting one-dimension channels by the application of  normal mode transformation (\ref{eq:modetrans}). The mode decomposition can be applied to a broader set of Hamiltonians than defined in (\ref{eq:Hd}). If we relax the condition of homogeneous frequencies and couplings to the condition of having them equal in all the dimensions except for the direction of transport, then the inhomogeneous system can be decomposed into one-dimension disordered chains. These disordered one-dimension systems have been broadly studied and are known to display numerical evidence of  non-ballistic behaviour \cite{bermudez:prl13,sun:epl10}.

\section{Spin lattices}
In one dimensional spin systems driven out of equilibrium by thermal baths (XY model) the transport is  similar to   a homogeneous fermionic chain. In the spin case the current is given by an expression similar to (\ref{eq:current}) \cite{manzano:pre12}. This is a natural result as one dimensional spins systems can be mapped locally to fermions by the Jordan-Wigner transformation \cite{jordan:28}. Spin chains with different Hamiltonians have been broadly studied \cite{manzano:pre12,sun:epl10,znidaric:jsm10}, concluding that in most of the cases the transport in a homogeneous and noise-free spin chain is ballistic and it becomes diffusive in the presence of static or dynamical noise.

On the other hand, the similarity between spin and fermionic systems breaks down if the dimensionality is larger than one. As we have proven, fermionic lattices are ballistic in any dimension as long as the couplings are homogeneous in each dimension. But it has been demonstrated that this behaviour does not hold for spin lattices. Even the simplest two-dimension spin lattice, a homogeneous ladder, presents both ballistic and non-ballistic invariant subspaces.  The relative size of the ballistic subspaces goes to zero as the size of the ladder goes to infinity \cite{znidaric:prl13}. This fact highlights the difference between spins and fermions as well as the importance of the dimensionality in spin transport. 

A general spin lattice of dimension $d$ has a Hamiltonian similar to the fermionic lattice (\ref{eq:Hd}). It can be decomposed in the form $H^{\text{spins}}=H^\text{spins}_{\text{free}}+\sum_{i=1}^d  H^\text{spins}_i$ with

\ben
H^\text{spins}_{\text{free}}&=&  \omega \sum_{\vec{l}}  \sigma^+_{\vec{l}} \sigma^-_{\vec{l}}  \nonumber\\
H^\text{spins}_i &=& g_i \sum_{\vec{l}} \pare{ \sigma^+_{l_1,\cdots,l_i,\cdots,l_d} \sigma^-_{l_1,\cdots,l_i+1,\cdots,l_d}  + \text{H.c.} },
\label{eq:Hd_spins}
\een
where $\sigma^+/\sigma^-$ are the Pauli raising/lowering operators. 

Due to the similarity of Hamiltonians (\ref{eq:Hd}) and (\ref{eq:Hd_spins}) it is natural to think that the mode transformation (\ref{eq:modetrans}) should also decompose a $d$-dimension spin lattice into independent one-dimension spin chains, but this is not the case. Both bosons and fermions have uniform commutation and anticommutation relations respectively, but spin systems have not \footnote{Note that if we have a spin system with sites labelled by the index $i$ we can define the dynamics by the Pauli operators $\key{\sigma_i^+,\sigma_i^-,\sigma_i^z}$. We can then define a set of {\it pseudofermionic} operators by stating $\sigma_i^+=f_i^\dagger$, $\sigma_i^-=f_i^{\phantom{\dagger}}$, and $\sigma_i^z=2f_i^\dagger f_i^{\phantom{\dagger}}-\mathbb{I}$. These operators fulfil the anticommutation relation $\key{f^\dagger_i,f^{\phantom{\dagger}}_i}=1$, but operators from different sites commute $\cor{f^\dagger_i,f^{\phantom{\dagger}}_j}=0$ ($\forall i\neq j$). There is no local transformation that transforms the Pauli operators into fermionic operators.}. Therefore, by applying the transformation (\ref{eq:modetrans}) to a $d$-dimension spin lattice, one-dimension mode operators can be defined. But these operators have different commutation rules compared to the original operators in the site basis. Because of that, this transformation does not map a $d$-dimension spin lattice into independent one-dimension spin chains.

Nevertheless, spin systems can be transformed to fermionic systems by the Jordan-Wigner transformation \cite{jordan:28}. This transformation is defined by

\be
f^\dagger_i =  \sigma^+_i \prod_{j<i} \sigma^z_j  , \quad 
f^{\phantom{\dagger}}_i = \sigma^-_i   \prod_{j<i} \sigma^z_j,   
\label{eq:jw}
\ee
where the index $i$ refers to an arbitrary ordering of the spins. Owing to this arbitrariness there is not a unique representation of the Jordan-Wigner transformation. In one-dimension systems with next-neighbours couplings this transformation can be performed in such a way that a homogeneous spin chain is mapped into a homogeneous fermionic chain. This explains why these two systems have the same behaviour. On the other hand, in systems with more than one dimension this is not true. Due to the non-local character of the Jordan-Wigner transformation, when it is applied to a multidimensional spin Hamiltonian the corresponding fermionic Hamiltonian includes also non-local terms.

\begin{figure}[h]
\includegraphics[scale=0.3]{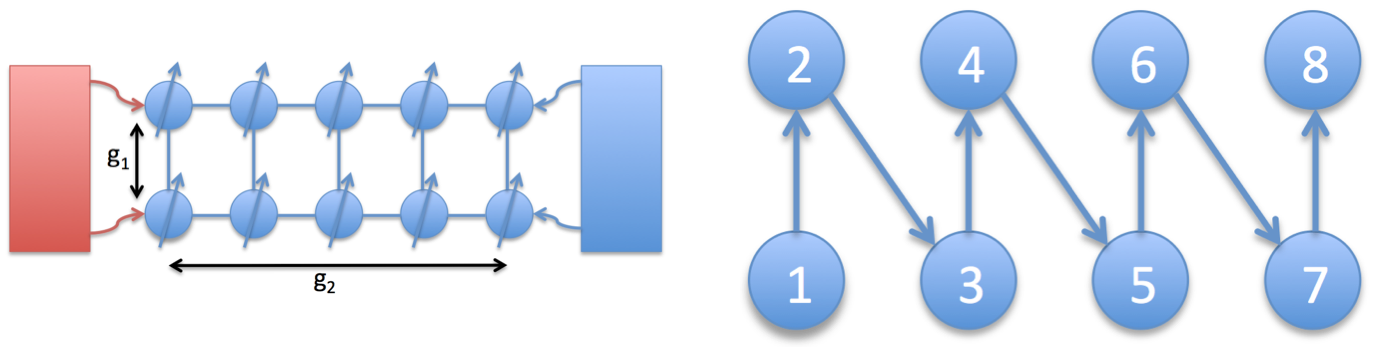}
\caption{(Left) Spin ladder coupled to two thermal baths at different temperatures. (Right) Labelling of the spin ladder.}
\label{fig:ladder}
\end{figure}

By the use of this transformation, we analyse the simplest spin lattice beyond one-dimension, a ladder. We order the sites by defining a new index $l=l_1 + 2 (l_2-1)$, where $l_1=1,2$ and $l_2=1,\dots,L$, and $L$ is the size of the system in the direction of the energy transfer (see figure \ref{fig:ladder}). The Hamiltonian of a uniform spin ladder in terms of the Pauli matrices is

\be
H^\text{spins}_{\text{ladder}} =  \omega \sum_{l=1}^{2L} \sigma_l^+ \sigma_l^-
+ g_1 \sum_{l=1}^{L}   \pare{\sigma_{2l}^+ \sigma_{2l-1}^- + \text{H.c.}}
+ g_2 \sum_{l=1}^{2L-2} \pare{\sigma_l^+ \sigma_{l+2}^-  + \text{H.c} }.
\label{eq:ladder_spin}
\ee
After applying the Jordan-Wigner transformation (\ref{eq:jw}) we obtain the fermionic Hamiltonian 

\be
H_{\text{ladder}} = \omega \sum_{l=1}^{2L} f_l^\dagger f_l^{\phantom{\dagger}} 
+ g_1 \sum_{l=1}^{L} \pare{f_{2l}^\dagger f_{2l-1}^{\phantom{\dagger}}+ \text{H.c.}}
+ g_2  \sum_{l=1}^{2L-2} \pare{ f_l^\dagger f^{\phantom{\dagger}}_{l+2} + \text{H.c.} }  N_{l+1},
\label{eq:ladder_fermion}
\ee
where $N_{i}=f^\dagger_{i}f^{\phantom{\dagger}}_{i} - f^{\phantom{\dagger}}_{i} f^\dagger_{i}$. This expression is similar to the fermionic ladder given by (\ref{eq:Hd}), but it includes a new term that is responsible for the non-ballistic behaviour already found in this system. On the other hand, as stated in Ref. \cite{znidaric:prl13} this system has a plethora of invariant subspaces. By direct analysis of Hamiltonian (\ref{eq:ladder_fermion}) the transport properties of each subspace can be inferred. If any state within a subspace, $\ket{k}$, fulfils that $N_{l+1} \ket{k}=c\ket{k}$  with a constant value of $c$ for those values of $l$ that satisfy $ f_l^\dagger f^{\phantom{\dagger}}_{l+2}\ket{k}\neq 0$ or $ f_{l+2}^\dagger f^{\phantom{\dagger}}_l \ket{k}\neq 0$, the system reduces to a uniform fermion ladder and the transport is therefore ballistic.

The role of the number of excitations in the non-ballistic character of the system is clear from (\ref{eq:ladder_fermion}). If the system contains only one excitation it is equivalent to a uniform fermion ladder, as in this case, and the transport is ballistic. This result can be extended straightforwardly to a more general case of a  $d$-dimension lattice with arbitrary size, proving that there can be a ballistic spreading of the excitation in all directions and extending previous results about discrete quantum walks in high dimensions \cite{krovi:pra06,makmal:pra14,patel:pra10} to the continuous domain.

\subsection{Examples}
   
\subsubsection{ Invariant subspace}
  
In Ref. \cite{znidaric:prl13} \v{Z}nidari\v{c} designed an open spin ladder with ballistic transport. This design is based on the use of non-local Lindblad superoperators that keep the system in a ballistic subspace. Together with a proper initial state he proved by numerical simulation that the flux is independent of the system size. The analysis of this system is simplified by using a Bell-type basis in the ladder rungs. This basis is defined by the vectors

\ben
  \ket{S}_i &=& \frac{1}{\sqrt{2}} \pare{ \sigma_i^+ + \sigma_{i+1}^+ }\ket{\text{vac}}_{i,i+1}, \nonumber\\
  \ket{T}_{i} &=& \frac{1}{\sqrt{2}} \pare{ \sigma_i^+ - \sigma_{i+1}^+ }\ket{\text{vac}}_{i,i+1},  \nonumber\\
  \ket{O}_{i} &=& \ket{\text{vac}}_{i,i+1}, \nonumber\\
  \ket{I}_i &=& \sigma_i^+ \sigma_{i+1}^+ \ket{\text{vac}}_{i,i+1},
 \een
 with $i$ being an odd number and $\ket{\text{vac}}_{i,i+1}$ being the vacuum state of sites $i$ and $i+1$. 
 
 The initial state of the system is composed by alternating entangled states in the form $\ket{STST\dots}$ or $\ket{TSTS\dots}$, and the Lindblad superoperators act non-locally
 
 \ben
 L_1 &=& \op{ITS\dots ST}{STS\dots ST},  \nonumber \\
 L_2 &=& \op{IST\dots TS}{TST\dots TS},  \nonumber \\
 L_3 &=& \op{STS\dots ST}{STS\dots SI},  \nonumber \\
 L_4 &=& \op{ST\dots TS}{TSTS\dots TI}.
 \een
   
The effect of the baths in this case is given by the superoperator   
 
\be
\mathcal{L}\rho= \sum_{i=1}^4 L_i^{\phantom{\dagger}}  \rho  L_i^\dagger - \frac{1}{2} \key{L_i^\dagger L_i^{\phantom{\dagger}},\rho}.
\ee 
We have not specified temperature or coupling strength for simplicity, as the ballistic character of the system is independent of these parameters. By applying the Jordan-Wigner transformation (\ref{eq:jw}) the Bell basis elements can be written as a function of  fermionic operators. 

\ben
 \ket{S}_i &=& \frac{1}{\sqrt{2}} \pare{ f_i^{\dagger} + f_{i+1}^{\dagger} }\ket{\text{vac}}_{i,i+1}, \nonumber\\
 \ket{T}_i &=& \frac{1}{\sqrt{2}} \pare{ f_i^{\dagger} - f_{i+1}^{\dagger} }\ket{\text{vac}}_{i,i+1}, \nonumber\\
 \ket{O}_{i} &=& \ket{\text{vac}}_{i,i+1}, \nonumber\\
 \ket{I}_i &=& f_i^\dagger f_{i+1}^\dagger \ket{\text{vac}}_{i,i+1}.
\een
The ballistic character of this subspace can be proved by direct inspection. All the states in the form $\ket{ST}_{i,i+1}$ (or $\ket{TS}_{i,i+1}$) are eigenvectors of the Hamiltonian and they are therefore invariant. Furthermore, for all the states $\ket{k}$ in the form $\ket{SI}_{i,i+1}$ (or $\ket{IS}_{i,i+1},\;\ket{TI}_{i,i+1},\;\ket{IT}_{i,i+1}$) with $ f_l^\dagger f^{\phantom{\dagger}}_{l+2}\ket{k}\neq 0$ or $ f_{l+2}^\dagger f^{\phantom{\dagger}}_l \ket{k}\neq 0$ one obtains $N_{l+1}\ket{SI}=\ket{SI}$ due to the entangled state of the background. Consequently, this subspace is ballistic, as it can be mapped into a homogeneous fermion system.

\subsubsection{Double exciton subspace}   
  
\begin{figure}[h]
\includegraphics[scale=0.6]{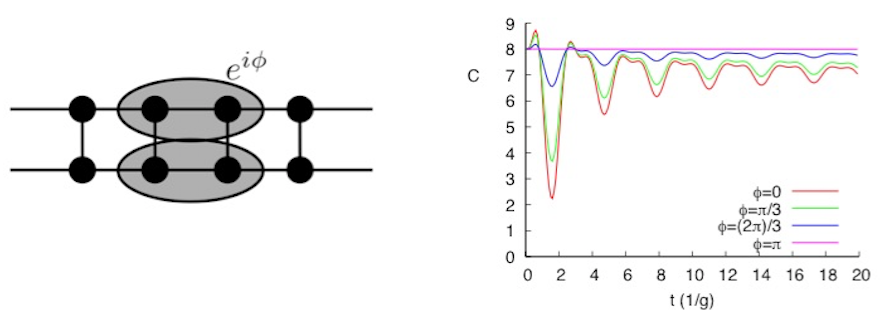}
\caption{Left: sketch of the initial state (\ref{eq:initial_msd}). Right: C for a ladder as a function of time for different values of $\phi$. The initial state is given by  (\ref{eq:initial_msd}).}
\label{fig:msd}
\end{figure}

In the absence of hot  and cold baths, if a system is prepared in an initial state that belongs to a ballistic subspace it will propagate ballistically without leaving the subspace. To study this behavour we have simulated the dynamics of a closed ladder. This allows us to analyze different initial conditions in a way that is impossible with a system coupled to baths. First, we study the dynamics of a spin ladder with $L=130$ and $g_1=g_2=g$ with an initial state given by 
\be
\ket{\psi(0)}_\phi=\pare{\sigma^+_{L-1} \sigma^+_{L+1}+ e^{i\phi} \sigma^+_{L} \sigma^+_{L+2}}\ket{vac},
\label{eq:initial_msd}
\ee
which is  a superposition of two excitations on the top and bottom legs at the middle of the ladde as it is displayed in figure 4 (left)r. To analyze the transport properties we have calculated the second derivative of the mean square displacement of the excitations in the direction of the energy transfer, given by $C=\frac{d^2\mean{x_2^2}}{dt^2}$. This is a common measure of transport properties \cite{msd}. For ballistic transport we have a constant value $C=4ng^2$, where $n$ is the total number of excitations in the system. The results are displayed in figure (\ref{fig:msd}) for different values of $\phi$. Only for $\phi=\pi$ the system is in an invariant subspace corresponding to a uniform fermion ladder and the transport is ballistic. This prediction is confirmed in figure \ref{fig:msd} right by the constant value of $C=8$. The opposite behaviour is found with $\phi=0$, and it is characterized by strong oscillations of $C$.

\subsubsection{Four exciton subspace}   

\begin{figure}[h]
\includegraphics[scale=0.5]{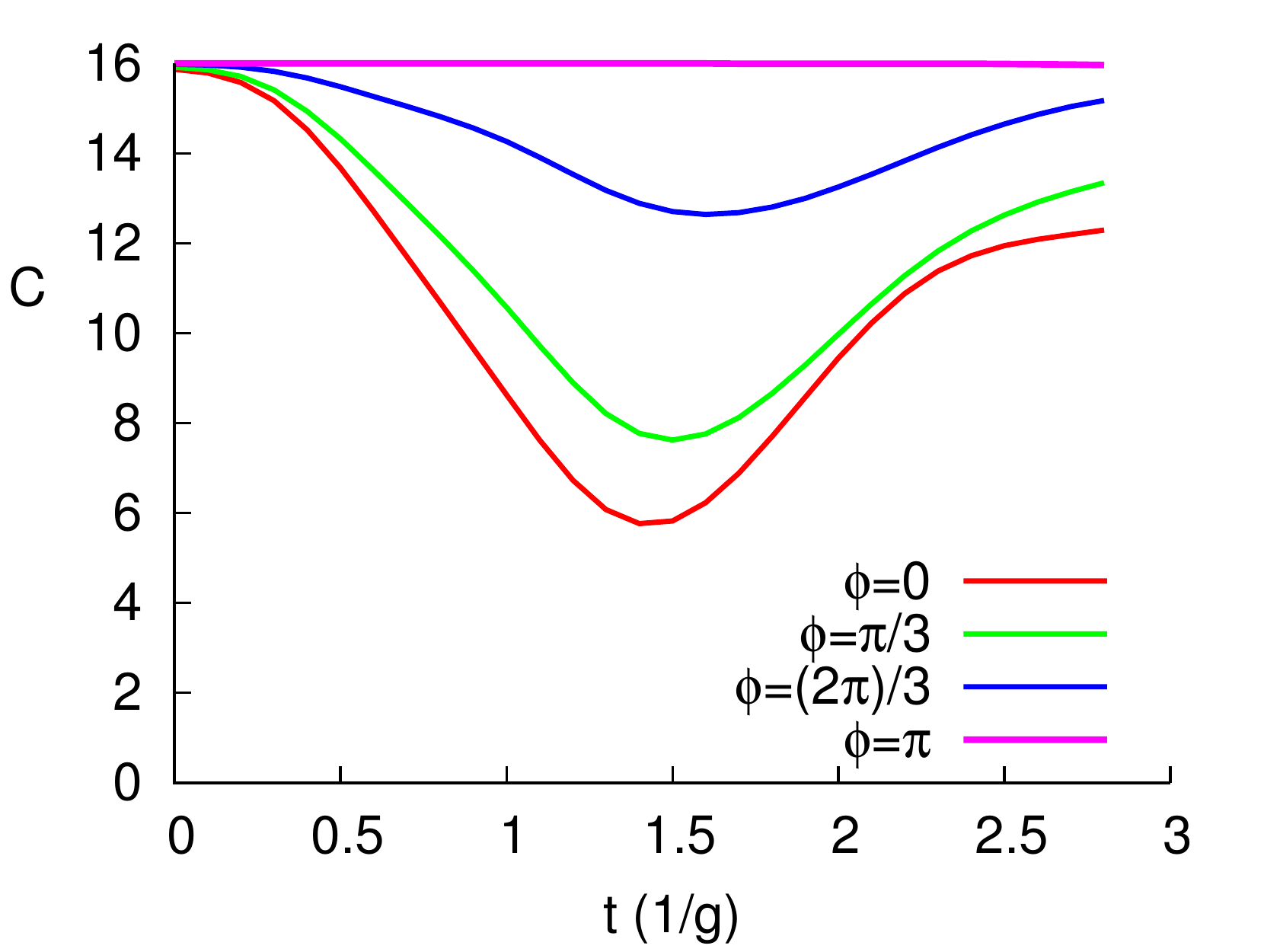}
\caption{C(t) for a ladder as a function of time. The initial state is given by (\ref{eq:initial_msd4}) with four excitations in the center of the ladder. We follow the same convention used in figure (\ref{fig:msd}).}
\label{fig:msd4}
\end{figure}

We provide another example inspired by Ref. \cite{znidaric:prl13}. We first define the Bell-type state
\be
\ket{A_\phi}_i=\frac{1}{\sqrt{2}}\pare{\sigma^+_i+e^{i\phi}\sigma^-_{i+1}}\ket{vac}_{i,i+1}.
\ee
It is clear that $\ket{\phi=0}=\ket{S}$ and $\ket{\phi=\pi}=\ket{T}$. The initial state is defined as
\be
\ket{\psi(0)}_\phi=\ket{O\dots OSA_\phi SA_\phi O\dots O},
\label{eq:initial_msd4}
\ee
where only the central sites are excited to alternating $S$ and $A_\phi$ states. With $\phi=\pi$ one restores the "$S-T$ background" introduced in Ref. \cite{znidaric:prl13}, where all other sites are in the ground state. The results with four excitations are shown in figure (\ref{fig:msd4}). Similar to the previous example, it is proven that with $\phi=\pi$ the system stays in the ballistic subspace indefinitely. This can also be generalised to states with arbitrarily higher or odd numbers of excitations. For example, the state $\ket{\psi(0)}=\ket{O\dots OSTSO\dots O}$ containing three excitations belongs to a ballistic subspace.

All the ballistic subspaces of the spin ladder described in Ref. \cite{znidaric:prl13} correspond to uniform fermion ladders. This result can be easily extended to general lattices, proving the existence of invariant ballistic subspaces for a spin lattice of an arbitrary size and dimension. For an open system the transport properties depend on the initial state and on the explicit form of the Lindblad superoperators that model the interaction with the baths, so that it is possible to design systems with multiple steady states with different currents \cite{manzano:prb14}. This allows the design of a spin lattice with ballistic and non-ballistic steady states and the control of the system current by regulating the symmetries of the system.

\section{\bf Conclusions} 

We have presented in this paper a normal  mode decomposition scheme that reduces $d$-dimensional bosonic and fermionic uniform lattices of arbitrary dimension into independent one dimension chains. These lattices are relevant for the study of many realistic systems such as optical lattices and cubic crystal lattices. By the use of the normal mode technique, the analytical expression for the energy transfer in an arbitrary uniform lattice is derived. For any dimension and size of the lattice, and with local heat baths acting at the ends of one of the dimensions, the system behaves ballistically and the energy transfer flux is independent of the system size. This result holds for both fermionic and bosonic systems. However, the dependence of the energy transfer on the bath temperature is different for fermionic and bosonic lattices, especially at high temperatures.

This method can also be extended to spin lattices, which has been previously proven as non-ballistic. The analysis is carried out by applying the Jordan-Wigner transformation, that transforms a spin system into a fermionic one. The application of mode-decomposition to spin lattices proves that in general a uniform spin lattice corresponds to a non-uniform fermionic lattice. Spin lattices have many invariant subspaces due to the high symmetry on the system topology. Some of these subspaces are equivalent to fermionic uniform lattices and are consequently ballistic. The remaining subspaces are equivalent to non-uniform fermionic systems. This result generalises previous studies of spin ladders to systems with arbitrary size and dimension. 

Finally, by applying the mode-decomposition technique we  demonstrate that a single excitation in a $d$-dimension spin lattice always propagates ballistically. This observation is relevant for the development of algorithms that use continuous quantum walks in a hypercubic lattice.  

\section{Acknowledgments}
This work was supported by the NSF (Grant No. CHE-1112825) and the center for Excitonics, an Energy Frontier Research Center funded by the US Department of Energy, Office of Science, Office of Basic Energy Sciences under Award No. DE-SC0001088. DM also acknowledges the Singapore University of Technology and Design for its support under the program MIT-SUTD.

\bibliographystyle{unsrt}

\newpage
\appendix*

\section{Appendix A. Calculation of the energy transfer in a fermionic chain}

The calculation of the energy transfer in a one-dimensional fermionic chain can be performed by applying the same method used in Ref. \cite{manzano:pre12}. The one dimensional system Hamiltonian reads 

\be
H=\omega \sum_{i=1}^L f_i^\dagger f_i^{\phantom{\dagger}} + g \sum_{i=1}^{L-1}\pare{f_i^\dagger f_{i+1}^{\phantom{\dagger}} + f_{i+1}^\dagger f_i^{\phantom{\dagger}}}, 
\ee
were $f_i^\dagger/f_i^{\phantom{\dagger}}$ are the creation/anhilitation fermionic operators, $L$ is the number of sites of the system and $\hbar=1$.  

The system is driven out of equilibrium by bosonic heat baths connected to the terminal qubits and describe by the Lindblad terms 

\ben
\mathcal{L}_{1/L} \rho &=& \gamma_{1/L} s_{1/L} \pare{ f_{1/L}^\dagger \rho f_{1/L}^{\phantom{\dagger}} - \frac{1}{2} \key{ f_{1/L}^{\phantom{\dagger}} f_{1/L}^\dagger,\rho }} 
\nonumber\\
&=& \gamma_{1/L} (s_{1/L}-1) \pare{ f_{1/L}^{\phantom{\dagger}} \rho f_{1/L}^\dagger  - \frac{1}{2} \key{ f_{1/L}^\dagger f_{1/L}^{\phantom{\dagger}},\rho }}.
\nonumber\\
\een

The full dynamics of the system is given by the equation 

\be
\dot{\rho}=-i \cor{H,\rho} + \mathcal{L}_1 \rho+ \mathcal{L}_L \rho.
\label{eq:me_1dim}
\ee

The energy flux is given by 

\ben
J&=&\Tr(H\mathcal{L}_1\rho) \nonumber\\
&=&\gamma_1 \pare{s_1-\esp{f_1^\dagger f_1^{\phantom{\dagger}}}} -\frac{\gamma_1 g}{2} \pare{\esp{f_1^\dagger f_2}^{\phantom{\dagger}} -\esp{f_1^{\phantom{\dagger}} f_2^\dagger}},  \nonumber\\
\label{eq:ef}
\een
with $\gamma_i=\Gamma_i(2 n_i + 1)$ and $s_i=n_i/(2n_i+1)$. The heat current at the steady state can be calculated only from the excited state population of the first site and the imaginary part of the coherence between this site and the second one. Due to the structure of the master equation (\ref{eq:me_1dim}) the next-neighbouring sites coherences are purely real, and the energy flux depends only in the first site population (see Ref. \cite{manzano:pre12}). Therefore, the problem of calculating the energy flux is reduced to the problem of calculating the population of the first site at the steady state. 

At the steady state the time-derivative of each expectation value is equal to zero. By calculating the factors $\frac{d}{dt} \esp{f_k^\dagger f_k^{\phantom{\dagger}}}=0$, with $k=1,L$ we obtain the pair of equations 

\ben
\gamma_1 \pare{s_1 - \esp{f_1^\dagger f_1^{\phantom{\dagger}}}}& =& -g \pare{  \esp{f_1^\dagger f_2^{\phantom{\dagger}}} + \esp{f_1^{\phantom{\dagger}} f_2^\dagger}} 
\nonumber\\
\gamma_L \pare{s_L - \esp{f_L^\dagger f_L^{\phantom{\dagger}}}}& =& g \pare{  \esp{f_{L-1}^\dagger f_L^{\phantom{\dagger}}} + \esp{f_{L-1}^{\phantom{\dagger}} f_L^\dagger}},
\nonumber\\
\label{eq:pair}
\een
for the terminal sites. Whereas for the inner sites the result is 
\be
\esp{f_{k-1}^\dagger f_k^{\phantom{\dagger}}} + \esp{f_{k-1}^{\phantom{\dagger}} f_k^{\dagger}} =
\esp{f_{k}^\dagger f_{k+1}^{\phantom{\dagger}}} + \esp{f_{k}^{\phantom{\dagger}} f_{k+1}^{\dagger}}.
\ee

The last equation is obtained by summing up the coherences, 
$\frac{d}{dt} \sum_{k=1}^{L-1} \esp{f_k^\dagger f_{k+1}^{\phantom{\dagger}}}=0$.

\be
\frac{\gamma_1}{2} \esp{f_1^\dagger f_2^{\phantom{\dagger}}} + \frac{\gamma_L}{2} \esp{f_{L-1}^\dagger f_L^{\phantom{\dagger}}}=
-ig \pare{ \esp{f_1^\dagger f_1^{\phantom{\dagger}}}  +  \esp{f_L^\dagger f_L^{\phantom{\dagger}}}}
\ee

By solving these five equations, we can calculate the first site population $\esp{f_1^\dagger f_1^{\phantom{\dagger}}}$, and by substituting in the energy flux equation (\ref{eq:ef}) we obtain

\be
J= \omega \frac{4g^2 \gamma_1\gamma_2 (s_1-s_2)}{\pare{\gamma_1+\gamma_2}
 \pare{4g_d^2+\gamma_1 \gamma_2} }
\ee

\end{document}